\documentclass[11pt, letter]{article}
\usepackage[utf8]{inputenc}
\usepackage[final]{microtype}

\usepackage[hyphens]{url}
\usepackage[square,numbers]{natbib}
\usepackage{authblk}

\usepackage{hyperref}
\usepackage{breakurl}

\usepackage{caption}
\usepackage{subcaption}

\usepackage{array,booktabs,longtable,tabularx}

\usepackage{amsmath,amsfonts,amssymb,amsthm}

\usepackage{mathtools}
\DeclarePairedDelimiter\floor{\lfloor}{\rfloor}

\usepackage{listings}
\usepackage[newfloat,frozencache,cachedir=.]{minted}

\newcommand{\code}[1]{``\lstinline[language=Python]!#1!''}
\newcommand{\codeq}[1]{\lstinline[language=Python]!#1!}

\usepackage[ruled,vlined]{algorithm2e}
\usepackage[inline]{enumitem}

\usepackage{diagbox}

\theoremstyle{definition}
\newtheorem{exmp}{Example}[section]

\usepackage{mathtools}
 
\DeclarePairedDelimiter{\abs}{\lvert}{\rvert}
\let\ket\relax
\DeclarePairedDelimiter{\ket}{\lvert}{\rangle}

\let\braket\relax
\DeclarePairedDelimiterX{\braket}[2]{\langle}{\rangle}{#1 \delimsize\vert #2}

\newcommand{\npat}{r}
\newcommand{\XOR}{\textsc{cnot}}

\begin{document}
\title{String Comparison on a Quantum Computer Using~Hamming~Distance}

\author[]{Mushahid Khan}

\author[]{Andriy Miranskyy}

\affil[]{Department of Computer Science, Ryerson University \protect\\ Toronto, Canada}
\affil[]{{\{mushahid.khan, avm\}@ryerson.ca}}

\date{}
\maketitle

\begin{abstract}
The Hamming distance is ubiquitous in computing. Its computation gets expensive when one needs to compare a string against many strings. Quantum computers (QCs) may speed up the comparison. 

In this paper, we extend an existing algorithm for computing the Hamming distance. The extension can compare strings with symbols drawn from an arbitrary-long alphabet (which the original algorithm could not). We implement our extended algorithm using the QisKit framework to be executed by a programmer without the knowledge of a QC (the code is publicly available). We then provide four pedagogical examples: two from the field of bioinformatics and two from the field of software engineering. We finish by discussing resource requirements and the time horizon of the QCs becoming practical for string comparison.

\end{abstract}

\maketitle

\section{Introduction}\label{sec:intro}
The Hamming distance~\cite{hamming1950error}, deemed $D$, computes the number of positions in which differences exist between two strings of equal length. For example, the distance between strings \code{010} and \code{101} is $D=3$; and the distance between strings \code{week} and \code{weak} is $D=1$.

The Hamming distance is ubiquitous and used in domains ranging from genetics~\cite{deaton1996good,mohammadi2017hamming} to cryptography~\cite{davida1999relation,shan2017machine}. In bioinformatics (BI), it is used for DNA sequence comparison~\cite{mohammadi2017hamming,sarkar2019algorithm}. It is also omnipresent in software engineering (SE), appearing in areas of static and dynamic analysis, such as test selection and generation~\cite{malaiya1995antirandom,wu2008antirandom,mrozek2012antirandom,hridoy2015regression}, code coverage inspection~\cite{hridoy2015regression,horvath2019code}, log or trace analysis~\cite{du2004anomaly}, and cybersecurity~\cite{du2004anomaly,taheri2020similarity,kanuparthi2016controlling,dario2017detecting}. The authors of these papers represent a collection of SE artifacts as symbols in a string. 

Computation of $D$ is inexpensive for a pair of strings of length $n$, taking $O(n)$ computations on a modern computer. However, when one needs to do it repetitively, say $r$ times, this becomes laborious~\cite{tang2015efficient}, as the computational complexity grows to $O(rn)$. Repetitive comparison frequently arises in bioinformatics, e.g., when a target DNA sequence needs to be compared against a database of sequences~\cite{mount2004bioinformatics}. This is also true for SE, e.g., when one wants to compare a new software trace to a set of existing traces~\cite{du2004anomaly,miranskyy2007iterative,miranskyy2008sift} or compare code coverage of a new test case to code coverage of existing test cases~\cite{hridoy2015regression}.

How can we speed up the process of comparison? Quantum computers (QCs) may come to the rescue. Trugenberger~\cite{pqmct} came up with an algorithm that computes the distance from a binary string to a group of strings on a QC to match bit strings. The string comparison is made at the bit level. The computational complexity of loading the strings into QC is $O(rn)$, and the complexity of  computing $D$ between the target string and $r$ strings is only $O(n)$. Thus, QC is more efficient than a classical computer at the comparison phase, making it advantageous for a large value of $r$~\cite{sousa2020parametric}.

To speed up computation of $D$, we can use~\cite{pqmct}. However, the string comparison in~\cite{pqmct} is made at the bit level rather than at the symbol level. 
Computing $D$ at the level of the symbol level rather than bit level is essential, as it may yield a different ranking of the strings, as shown in the example below.

\begin{exmp}
Consider a target string $s_t =$~``\texttt{00~00~00}'', where symbols are codified using two bits. That is, our target string consists of three consecutive symbols ``\texttt{00}''. We want to compute the distance between $s_t$ and two strings: $s_1$ = ``\texttt{01 01 01}'' and $s_2$ = ``\texttt{11 11 00}''. At the bit level, $s_1$ is closer to $s_t$ than $s_2$, as $D_{\textrm{bit}}(s_t, s_1) = 3$ and $D_{\textrm{bit}}(s_t, s_2) = 4$. At the symbol level, the ranking is the opposite, as  $D_{\textrm{symbol}}(s_t, s_1) = 3$ and $D_{\textrm{symbol}}(s_t, s_2) = 2$.
\end{exmp}

This paper will extend the core ideas of~\cite{pqmct} to compute $D$ for symbols codified by multiple bits. To do this, we extend\footnote{The code is made available via GitHub~\cite{dat:github}.} the implementation of~\cite{pqmct} for a modern ``noisy'' QC~\cite{sousa2020parametric}. 
We will also show pedagogical examples of using these approaches and discuss practical applicability.

\section{Prior Art}
\subsection{Literature Review}
\subsubsection{Usage of QC algorithms in BI and SE}
To the best of our knowledge, there are no QC algorithms that were applied to SE-specific problems. In BI, QC algorithms involving  $D$ are used to classify individuals with disease versus control~\cite{kathuria2020implementation} and to approximate pattern-matching for DNA read alignment~\cite{sarkar2019algorithm}. In~\cite{kathuria2020implementation,sarkar2019algorithm}, $D$ is calculated at the bit-level instead of at the symbol-level. Thus, these works are complementary to ours.

\subsubsection{Computation of $D$ on a QC}
Let us now explore how $D$ can be computed using a QC.
There has been work done in the field of QC for calculating $D$ between two binary strings at the bit level. In~\cite{bravo2018calculating, doriguello2019quantum}, an algorithm is presented to calculate the $D$ of two binary strings of equal length. 
In~\cite{hamdistancebooleanfunctions2018}, $D$ between two $n$-variable boolean functions is calculated. 
In all of these works, the pair-wise comparison is implemented. However, it cannot be extended to efficiently compare a single string against a group of strings. This is where Probabilistic Quantum Memory (PQM)~\cite{pqmct} may become helpful, as we will discuss in Section~\ref{sec:pqm}. Before doing this, let us introduce core QC fundamentals needed to implement the PQM (see, e.g.,~\cite{nielsen_chuang_2010} for extra details of QC fundamentals). 

\subsection{Quantum Computing} \label{sec:qc}
In quantum computing, a qubit represents the basic unit of information. It is a two-level quantum system. Based on the State Space Postulate, it is described by a linear combination
$\ket{\psi} =  \alpha \ket{0} + \beta \ket{1} = \alpha \begin{bmatrix} 1 & 0 \end{bmatrix}^\intercal + \beta \begin{bmatrix} 0 & 1 \end{bmatrix}^\intercal$,
where $\alpha$ and $\beta$ are probabilistic amplitudes, such that  $\abs{\alpha}^2 + \abs{\beta}^2 = 1$. The state  $\ket{\psi}$ is said to be in a superposition of states $\ket{0}$ and $\ket{1}$. A qubit can be in more than one basis state at a given time. In the above, the qubit is in the computational basis.

To modify a state, quantum gates are used. A quantum gate is a basic quantum circuit that operates on a qubit. Quantum gates are the building blocks of quantum circuits. Quantum gates are unitary operators described as unitary matrices relative to a basis.  For the computational basis, the quantum gates that we will use in this paper are shown below.

Hadamard gate, given by 
$$\textsc{h} = \frac{1}{\sqrt{2}} \begin{bmatrix} 1 & 1\\ 1 & -1\\  \end{bmatrix}, $$
transforms a state into a superposition state. 
It maps  $\ket{0} \to \frac{1}{\sqrt{2}}  (\ket{0} + \ket{1})$ and $\ket{1} \to \frac{1}{\sqrt{2}}(\ket{0} - \ket{1})$. 

Pauli-X gate is defined as
$$\textsc{x} = \begin{bmatrix} 0 & 1\\ 1 & 0\\  \end{bmatrix}.$$
It is the \textsc{not} gate which flips the qubit, transforming the state $\ket{0}$ to $\ket{1}$ and the state $\ket{1}$ to $\ket{0}$. 

Controlled-NOT gate, given by
$$\textsc{cnot} = \begin{bmatrix} 1 & 0 & 0&0\\ 0 & 1 & 0&0\\ 0 & 0 & 0&1 \\ 0 & 0 & 1&0\end{bmatrix},$$
is a two-qubit gate. It applies the \textsc{x} gate to a target qubit whenever its control qubit is $\ket{1}$; in this context, we can interpret it as the classic \textsc{xor} gate. \textsc{cnot} gate can be extended to having $n$ control qubits (\textsc{c}\textsuperscript{n}\textsc{not}). In this case, the \textsc{x} gate is applied to the target qubit whenever each of the $n$ control qubits are $\ket{1}$.

When a quantum system is not measured, a qubit can be in both states of $\ket{0}$ and $\ket{1}$. However, after measurement, the system collapses into either the state $\ket{0}$  or $\ket{1}$ with a probability of the absolute value of the amplitude squared.

\subsection{Probabilistic Quantum Memory}\label{sec:pqm}
Let us now review PQM~\cite{pqmct}, a distance computing data structure. Essentially, it is a probabilistic model that uses $D$ between the target pattern (string) and all the stored patterns (strings) to compute matching. PQM is designed to have scalable storage capability. It can store all possible binary patterns for $n$ bits. PQM has two parts: storing information and retrieving information, discussed in Sections~\ref{sec:orig_storage} and~\ref{sec:orig_retrieval}, respectively. We extend the retrieving information part of the model in Section~\ref{sec:our_retrieval}.

\subsubsection{PQM: Storing Information} \label{sec:orig_storage}
The storing information part of the algorithm receives a dataset of $r$ binary patterns, each of $n$ bits. To store the patterns, three quantum registers are needed: $\ket{p}$, $\ket{u}$, and $\ket{m}$. $\ket{p}$ will hold every pattern of length $n$ by acting as the input register  before it is processed and stored on $\ket{m}$, the memory register. $\ket{u}$ is an auxiliary two-qubit register. It is used to keep tabs on which patterns are stored in memory and which ones need to be processed. The full initial quantum state  after pattern $p^k$ has been stored in the input register will be 
\begin{equation*}
\ket{\phi^k_0} = \ket{p^k_1 p^k_2 \ldots p^k_n;01;0_10_2\ldots0_n}.
\end{equation*}
The algorithm ends up storing each pattern on the memory register. This process is explained in depth in~\cite{sousa2020parametric}. We use the algorithm unchanged and, to avoid redundancy, refer the reader to~\cite{sousa2020parametric} for details. 

\subsubsection{PQM: Retrieving Information} \label{sec:orig_retrieval}
The retrieving information part of the algorithm requires a copy of the memory register used in the storing information algorithm. 

This algorithm uses $D$ between a target pattern and all patterns, which are stored in a superposition, to indicate probabilistically the chances of the target pattern being in the memory. This algorithm uses three quantum registers: $\ket{s}$, $\ket{m}$, and $\ket{c}$. The target pattern, deemed $t$, is loaded into register $\ket{s}$; $\ket{m}$ contains all the stored patterns from the storage algorithm; and $\ket{c}$ contains a control qubit initialized in a uniform superposition of the states $\ket{0}$ and $\ket{1}$. Once the input has been loaded on to $\ket{s}$ and the stored patterns from the storing part of the algorithm have been copied over to $\ket{m}$, the full initial quantum state is
\begin{equation*}
\begin{split}
\ket{\psi_0} &= \frac{1}{\sqrt{2\npat}} \sum_{k = 1}^{\npat} \ket{s_1 s_2 \ldots s_n;m_1^k m_2^k \ldots m_n^k;0}\\
&+ \frac{1}{\sqrt{2\npat}} \sum_{k = 1}^{\npat} \ket{s_1 s_2 \ldots s_n;m_1^k m_2^k \ldots m_n^k;1},
\end{split}
\end{equation*}
where $r$ is the total number of stored patterns, $s_1s_2\ldots s_n$ is the target pattern $t$, and $m_1^k m_2^k \ldots m_n^k$ is the $k$-th stored pattern.
\begin{algorithm}[t]
    1. $\ket{\psi_1 } = \prod_{j=1}^n X_{m_j} \XOR_{s_j, m_j} \ket{\psi_0 }$ \\
    2. $\ket{\psi_2 } = \prod_{e=1}^n \left(GU^{-2}\right)_{c,m_e} \prod_{j=1}^n U_{m_j} \ket{\psi_1 }$ \\
    3. $\ket{\psi_3 } = H_c\prod_{j=n}^1  \XOR_{s_j, m_j}X_{m_j} \ket{\psi_2 }$\\
    4. Measure qubit $\ket{c}$\\
    5. \If{c == 0}{
        Measure the memory to obtain the desired state.
    }
    6. End
    \caption{Retrieving information~\cite{sousa2020parametric}}
    \label{alg:recover_orig}
\end{algorithm}

The retrieval process is summarized in Algorithm~\ref{alg:recover_orig}. Step 1 sets the $j$-th qubit in register $\ket{m}$ to $\ket{1}$ if the $j$-th qubit of  $\ket{s}$ and $\ket{m}$ are the same or to $\ket{0}$ if they differ. 

Step 2 computes $D$ between the target pattern and all patterns in $\ket{m}$. The number of zeros in $\ket{m}$ (representing the qubits that differ between memory and target string) is computed.
Operator $U$, used in this step, is defined as
\begin{equation*}
U = \left[\begin{array}{cc}
\exp\left(\frac{i \pi}{2n}\right) &   0 \\
0                  &   1
\end{array}\right], 
\end{equation*}
where $i$ denotes unit imaginary number.  First, $U$ is applied to each qubit in  $\ket{m}$. Then $U^{-2}$ is applied to each qubit in $\ket{m}$ if the qubit in $\ket{c} = \ket{1}$. This if-condition is denoted by the operator $G$.

Step 3 reverts register $\ket{m}$ to its original state and $H$ is applied to the control qubit in $\ket{c}$; this operation is denoted by $H_c$. 

Step 4 measures register $\ket{c}$. A target pattern similar to the stored patterns increases the probability of measuring $\ket{c} = 0$. Otherwise, if the target pattern is dissimilar, then the probability of $\ket{c} = 1$ increases. If $\ket{c}$ is measured and the output is $0$, then measuring the qubits in the memory register (in Step 5) will return the binary pattern from the stored patterns that has minimum $D$ with the target pattern. 

We then need to execute the circuit $N$ times to get statistics for the measured samples. The more frequent an observed pattern is, the more probable it is.

\section{Computation of $D$ for symbols of arbitrary length}\label{sec:our_retrieval}
PQM outputs the probability of a target pattern being close to patterns in the database at the bit level.
As discussed in Section~\ref{sec:intro}, this approach is ineffective if we compute $D$ for symbols represented by multiple bits. For this, we extend the information retrieval part of PQM.

In~\cite{sousa2020parametric}, it was suggested to store register $\ket{s}$ in classical computer's space, thus, reducing the required number of qubits. They call this approach ``hybrid classical/quantum protocol''. We follow their lead. 

Our extension requires one more register, $\ket{h}$. The number of qubits in $\ket{h}$ is equal to the number of symbols in the pattern, deemed $z$. Note that $z = n/d$, where $d$ is the number of bits required to represent a symbol in the alphabet. With $\ket{h}$ included and the hybrid classical/quantum protocol implementation, the initial quantum state for retrieval of information is
\begin{equation}\label{eq:our_state}
\begin{split}
\ket{\psi_0} &= \frac{1}{\sqrt{2\npat}} \sum_{k = 1}^{\npat} \ket{m_1^k m_2^k \ldots m_n^k;0;h_1 h_2 \ldots h_z}\\
&+ \frac{1}{\sqrt{2\npat}} \sum_{k = 1}^{\npat} \ket{m_1^k m_2^k \ldots m_n^k;1;h_1 h_2 \ldots h_z}.
\end{split}
\end{equation}
Initially, all qubits in register $\ket{h}$ will be set to $\ket{1}$.

With this extension, the retrieval algorithm is changed, as shown in Algorithm~\ref{alg:recover_ours}. The first step is the same as the original one. In Step 2, results from step 1 are used to update register $\ket{h}$. Given that each symbol is represented by a binary string of length $d$ and each input is of length $n$, Step 2 will set the $j$-th qubit in $\ket{h}$ to $\ket{1}$ if the binary string of length $d$ that represent $j$-th symbol of $t$ is the same as the corresponding binary string in $\ket{m}$. With register $\ket{h}$ updated in Step 2, $\ket{h}$ can be used in place of $\ket{m}$ for Step 3. In Step 3, we use operator $W$ instead of $U$, which is defined as
\begin{equation*}
W = \left[\begin{array}{cc}
\exp\left(\frac{i \pi}{2z}\right) &   0 \\
0                  &   1
\end{array}\right].
\end{equation*}
We use $W$ to adjust for calculating $D$ at the symbol level. In Steps 4 and 5, inverse transformations of Steps 1 and 2 are applied and $H$ is applied to the control qubit. If $\ket{c}$ is measured and the output is $0$, then measuring the qubits in memory register will return the binary pattern from the stored pattern that has the minimum $D$ with the target pattern at the symbol level. Finally, Steps 6 and 7 are identical to Steps 4 and 5 of Algorithm~\ref{alg:recover_orig}.

\begin{algorithm}[t]
    1. $\ket{\psi_1 } = \prod_{j=1}^n X_{m_j} \XOR_{s_j, m_j} \ket{\psi_0 }$ \\
    2.  $\ket{\psi_2 } = \prod_{j=1}^z
    X_{h_j}\textsc{c}\textsuperscript{d}\textsc{not}_{m_{d  (j-1) + 1 } m_{d(j-1)+2}\ldots m_{d j}, h_j} \ket{\psi_1 }$ \\ 
    3. $\ket{\psi_3 } = \prod_{e=1}^z \left(GW^{-2}\right)_{c,h_e} \prod_{j=1}^z W_{h_j} \ket{\psi_2 }$ \\
    4. $\ket{\psi_2 } = \prod_{j=z}^1
    \textsc{c}\textsuperscript{d}\textsc{not}_{m_{d  (j-1) + 1 } m_{d(j-1)+2}\ldots m_{d j}, h_j} X_{h_j}  \ket{\psi_3 }$ \\ 
    5. $\ket{\psi_5 } = H_c\prod_{j=n}^1  \XOR_{s_j, m_j}X_{m_j} \ket{\psi_4 }$\\
    6. Measure qubit $\ket{c}$\\ 
    7. \If{c == 0}{
        Measure the memory to obtain the desired state.
    }
    8. End
    \caption{Information retrieval (our extension of Alg.~\ref{alg:recover_orig})}
    \label{alg:recover_ours}
\end{algorithm}

\subsection{Post-processing on classical computer}
Once we get $N$ measurements from the QC (as discussed in Section~\ref{sec:orig_retrieval}), we convert the frequency of occurrence into $p$-values by normalizing the number of observations by $N$. The closer the probability value to $1$ --- the lower the $D$ is. In~\cite[Eq. 19]{pqmct}, the relation between $p$-values and $D$ are given for binary strings:
\begin{equation*}
P\left(p^k\right) = \frac{1}{rc} \cos^2{\left[\frac{\pi}{2n}  D\left(t,p^k\right)\right]}.
\end{equation*}
We extend this relation to symbols of arbitrary length as follows:
\begin{equation}\label{eq:p_and_d}
P\left(p^k\right) = \frac{1}{rc} \cos^2{\left[\frac{\pi}{2z}  D\left(t,p^k\right)\right]},
\end{equation}
where $P(p^k)$ is the $p$-value of pattern $p^k$ (which we passed to QC in Section~\ref{sec:orig_storage}), $c$ is $P(\ket{c} =\ket{0})$ and $D(t, p^k)$ is the $D$ between target pattern $t$ and pattern $p^k$. Solving~Eq.~\ref{eq:p_and_d} for $D$, we get
\begin{equation}\label{eq:d_and_p}
 D\left(t,p^k\right) \approx \frac{z}{\pi} \arccos\left[2crP\left(p^k\right) - 1\right].
\end{equation}
$D$ is an integer, while $c$ and $P(p^k)$ are real numbers that may change slightly from run-to-run of a QC, hence the approximation in Eq.~\ref{eq:d_and_p}.

\section{Implementation}\label{sec:implementation}
To implement our algorithm, we use one circuit. This circuit contains both the information storage part of~\cite{pqmct} and the extension of information retrieval we came up with. This way, we do not have to rely on a probabilistic cloning machine for the memory register as it was done in~\cite{pqmct}, which required $2n + 2$ qubits for storage and $2n + 1$ qubits for retrieval, making a total of $4n + 3$ qubits. Also, the input registers used for storage and retrieval remain in a classical state~\cite{sousa2020parametric}. This hybrid classical/quantum protocol enabled us to remove the input quantum registers. As a result, we need a total of $n + n/d + 2$ qubits reused in the storage and retrieval phases. 

Our algorithm can be implemented on any modern QC architecture. We base a reference implementation on QisKit~\cite{Qiskit}, a Python-based open-source software development kit for coding in OpenQASM and leveraging the IBM QCs. The code is given in~\cite{dat:github}.

Similar to the QisKit Aqua~\cite{AquaAlgo12:online} library, we wrap OpenQASM invocations into a Python class so that a programmer without any QC coding experience can leverage the algorithm from any Python program. The code can be executed in a simulator on a personal computer or on the actual IBM QC.

The algorithm is invoked by instantiating\footnote{The class constructor takes an additional parameters with default values, namely the pointer to the backend on which we will execute the code; we omit standard backend initialization for the sake of brevity.} the class \code{StringComparator(target, db, is_binary, symbol_length, shots)}, where \code{db} is a list of strings that we want to compare to the \code{target} string. The number of characters that codify a symbol (default value is 1) is given by \code{symbol_length}. \code{is_binary} specifies if we are passing binary strings (default behaviour) or lists of symbols. \code{shots} specifies $N$, i.e., the number of times QC has to repeat the circuit (default value\footnote{We set it to 8192 to align with the current max value of shots on the modern IBM QC. This is not a hardware limitation; rather, it is set to simplify job scheduling. Note that we can get a higher number of shots on the IBM QC by executing the circuit multiple times and aggregating raw shots count. } is 8192). The larger the value of $N$~--- the more accurate the measurements are, which is typical for a QC. We then execute the circuit by invoking \code{run()} method of the class. The method returns a data structure containing various helpful information about the execution of the code, including a list of $D$ values given by the \code{hamming_distances} field.

\subsection{Software engineering usage examples}\label{sec:usage_examples}
Below, we provide two toy examples of using the algorithm from SE domain. The same principle can be used to apply the algorithm to the more elaborate use-cases. We execute the examples in QisKit's QC simulator. 

\subsubsection{Code coverage}\label{sec:code_cov}
Let us look at the SE use-case, which can be reduced to bit strings comparison, i.e., strings with symbols drawn from an alphabet of length two.

Consider the code coverage problem of finding a set of test cases closest to a given test case (e.g., to restructure regression test suite~\cite{hridoy2015regression}). We can mark each code block by a numeric id which will correspond to a specific element in a bit string. If the $i$-th code block is covered with a test case, we will set the $i$-th element in the bit string to \code{1} or keep it at \code{0} otherwise.

Suppose we have a program with five code blocks, i.e., we will represent them as a 5-bit string. Our target test case covers blocks 1, 3, and 4. We will encode this coverage as \code{10110}. Our four existing test cases and blocks that they cover are listed in Table~\ref{tbl:ex_cov}.

We call \code{StringComparator} using the code in Listing~\ref{list:cov} and obtain approximation of $D$ based on Eq.~\ref{eq:d_and_p} in return. The actual values of $D$, returned by the program, match the expected values. 

\begin{listing}[h]
\caption{Code coverage example}\label{list:cov}
\begin{minted}[xleftmargin = 4mm, fontsize = \footnotesize, numbersep = 2mm, linenos = true]{python3}
from string_comparison import StringComparator

target = '10110'
db =    ['10110', '11010', '01110', '01001']
x = StringComparator(target, db)
results = x.run()
print(f"D = {results['hamming_distances']}")

# Output:
# D = [0, 2, 2, 5]
\end{minted}
\end{listing}

\begin{table}[t]
\caption{``Databases'' for examples in Section~\ref{sec:usage_examples}}
\begin{subtable}{0.45\columnwidth}
\centering
\begin{tabular}{lr}
\toprule
Test case & $D$ \\ \midrule
\code{10110}            & 0   \\
\code{11010}            & 2   \\
\code{01110}            & 2   \\
\code{01001}            & 5   \\ \bottomrule
\end{tabular}
\caption{Target = \code{10110}}
\label{tbl:ex_cov}
\end{subtable}
\begin{subtable}{0.45\columnwidth}
\centering
\begin{tabular}{lr}
\toprule
Trace & $D$ \\ \midrule
\code{foo quux bar}   & 1   \\
\code{foo bar  foo}   & 1   \\
\code{bar foo  foo}   & 2   \\
\code{foo bar  bar}   & 2   \\  \bottomrule
\end{tabular}
\caption{Target = \code{foo quux foo}}
\label{tbl:ex_trace}
\end{subtable}%
\end{table}

\subsubsection{Trace comparison}\label{sec:trace_comparison}
Let us now consider the SE use-case that will require comparison of strings with symbols drawn from an alphabet of length three (i.e., the bit-string-level comparison is not sufficient).

Suppose that we capture an execution trace of software. We will then compare them to existing traces to find similarities (e.g., to detect a defect or a cyber attack~\cite{du2004anomaly}). We will need to partition traces into sub-traces of the identical length as per~\cite{du2004anomaly} to use our algorithm.

For simplicity, let us assume that we capture only entry points into the functions, i.e., trace points will be equivalent to function names. There are three unique functions in the software: \code{foo}, \code{bar}, and \code{quux}, which we can codify using 2-bit symbols \code{00}, \code{01}, and \code{10}. Suppose our target trace is \code{foo quaz foo} represented by a bit string \code{00 10 00}. 

We can pass these binary strings to \code{StringComparator} directly (we simply need to set \code{symbol_length=2}). However, it is more convenient to pass the strings  as lists of symbols and set the constructor's parameter  \code{is_binary=False}. Under the hood, we automatically create the alphabet, determine the minimum number of bits needed to represent each symbol, and map the symbols to their bit representations.

The list of existing traces and associated outputs are summarized in Table~\ref{tbl:ex_trace} and the associated code in Listing~\ref{list:trace}. As in the code coverage case, the actual $D$ values (returned by the program) matched the expected ones. 

\begin{listing}[h]
\caption{Trace comparison example}\label{list:trace}
\begin{minted}[xleftmargin = 4mm, fontsize = \footnotesize, numbersep = 2mm, linenos = true]{python3}
from string_comparison import StringComparator

target = ['foo', 'quux', 'foo']
db =    [['foo', 'quux', 'bar'],
         ['foo', 'bar',  'foo'],
         ['bar', 'foo',  'foo'],
         ['foo', 'bar',  'bar']]
x = StringComparator(target, db, is_binary=False)
results = x.run()
print(f"D = {results['hamming_distances']}")

# Output:
# D = [1, 1, 2, 2]

\end{minted}
\end{listing}

\subsection{Bioinformatics usage examples}\label{sec:bioinformatics}
In this section, we give two toy examples of using the algorithm from the BI domain. As above, the code is executed in QisKit's QC simulator. 

\begin{table}[t]
\caption{``Databases'' for examples in Section~\ref{sec:bioinformatics}}
\begin{subtable}{0.45\columnwidth}
\centering
\begin{tabular}{lr}
\toprule
DNA sequence & $D$ \\ \midrule
\code{C G A A T T}            & 0   \\
\code{C C A A C C}            & 3   \\
\code{G A A A G A}            & 4   \\
\code{C G A T A T}            & 2   \\ \bottomrule
\end{tabular}
\caption{Target = \code{C G A A T T}}
\label{tbl:ex_dna}
\end{subtable}
\begin{subtable}{0.45\columnwidth}
\centering
\begin{tabular}{lr}
\toprule
mRNA sequence & $D$ \\ \midrule
\code{AUG ACG CUU}   & 1   \\
\code{GAG CGC CCC}   & 2   \\
\code{AAA ACG UUU}   & 2 \\  
\code{AGA GAG UUU}   & 3 \\  

\bottomrule
\end{tabular}
\caption{Target = \code{AUG ACG CCC}}
\label{tbl:ex_mrna}
\end{subtable}%
\end{table}

\subsubsection{Nucleotide-level DNA comparison}
Suppose we would like to compare a DNA sequence against a database of sequences. Each symbol in the sequence represents a nucleic acid. There are four commonly found nucleotides in DNA: adenine, cytosine, guanine, and thymine, denoted by symbols \codeq{A}, \codeq{C}, \codeq{G}, and \codeq{T}, respectively. 
We want to compare a DNA sequence of length six against four other DNA sequences listed in Table~\ref{tbl:ex_dna}. Given that we have four unique symbols, each one should be encoded by two bits. Similar to the approach in Section~\ref{sec:trace_comparison}, we pass the sequences  as lists of symbols, shown in Listing~\ref{list:dna}.  As before, the actual $D$ values (returned by the program) matched the expected ones. Note that we need to increase the number of shots from the default value of 8192 to 10000 to obtain consistently correct results\footnote{As discussed in the beginning of Section~\ref{sec:implementation}, the larger the sample size --- the more robust the results are.}.

\begin{listing}[h]
\caption{DNA comparison example}\label{list:dna}
\begin{minted}[xleftmargin = 4mm, fontsize = \footnotesize, numbersep = 2mm, linenos = true]{python3}
from string_comparison import StringComparator

target = ['C', 'G', 'A', 'A', 'T', 'T']
db =    [['C', 'G', 'A', 'A', 'T', 'T'], 
         ['C', 'C', 'A', 'A', 'C', 'C'], 
         ['G', 'A', 'A', 'A', 'G', 'A'],  
         ['C', 'G', 'A', 'T', 'A', 'T']]
x = StringComparator(target, db, 
                     is_binary=False, shots=10000)
results = x.run()
print(f"D = {results['hamming_distances']}")

# Output:
# D = [0, 3, 4, 2]
\end{minted}
\end{listing}

\subsubsection{Codon-level mRNA comparison}
Suppose we would like to compare an mRNA sequence encoded by the codons against three other sequences, listed in Table~\ref{tbl:ex_mrna}. A codon is a sequence of three nucleotides. The associated code and the output are given in Listing~\ref{list:mrna}. The actual results match the expected ones. 

Note that there exist 64 possible codons. However, in our example, we need to represent only nine (i.e., each codon is encoded by four bits per symbol), as this is the number of distinct symbols in the target and database strings.

\begin{listing}[h]
\caption{mRNA comparison example}\label{list:mrna}
\begin{minted}[xleftmargin = 4mm, fontsize = \footnotesize, numbersep = 2mm, linenos = true]{python3}
from string_comparison import StringComparator

target = ['AUG', 'ACG', 'CCC']
db =    [['AUG', 'ACG', 'CUU'], 
         ['GAG', 'CGC', 'CCC'], 
         ['AAA', 'ACG', 'UUU'],
         ['AGA', 'GAG', 'UUU']]
x = StringComparator(target, db, is_binary=False)
results = x.run()
print(f"D = {results['hamming_distances']}")

# Output:
# D = [1, 2, 2, 3]
\end{minted}
\end{listing}

\section{Practical considerations}
\subsection{Hardware constraints}

To implement our algorithm, as mentioned in Section~\ref{sec:implementation}, we require
\begin{equation}\label{eq:registry_size}
    n+n/d+2
\end{equation}
qubits. Thus, our space complexity is $O(n)$. Note that space complexity does not depend on the number of strings $r$, as they all simultaneously reside in the same registry in the superposition state (which is the beauty of quantum computing).

To initialize the database, we need to alter the state of qubits responsible for storing stings’ symbols: the database initialization complexity is $O(rn)$. Theoretically, we can do this an infinite number of times. That is, the number of strings $r$ can be arbitrarily large. 

However, the properties of an actual QC will affect the quality of results~\cite{Preskill2018quantumcomputing}. Here are four examples.
\begin{enumerate}
    \item After a while, the qubits of a modern QC will become decoherent (i.e., spontaneously change their state). Thus, there will be a limit (different for various QCs) on how many strings can be loaded into the database due to time and noise constraints. 
    \item Our algorithm uses sequences of \textsc{cnot} gates, each one introducing noise into the system, which leads to measurement errors~\cite{Preskill2018quantumcomputing}. 
    \item In a modern QC, not all the qubits are interconnected. This requires the transpiler to add \textsc{swap} gates to map\footnote{This is an NP-complete problem~\cite{10.1145/3168822}.} logical qubits to physical qubits, further amplifying the noise~\cite{li2019tackling}. 
    \item Right now, we have to re-initialize the databases for every new target string. Theoretically, this overhead can be reduced when partial measurements will be introduced to QC architectures~\cite{QiskitPartialMeasurement}. 
\end{enumerate}
These are hardware limitations that engineers will alleviate in the future as QCs evolve.

\subsection{Timeline}
When will we be able to use our algorithm for practical-scale applications? By 2023, IBM promises to ship an $1121$-qubit QC~\cite{cho2020ibm}. What can we readily use this machine for? Based on Eq.~\ref{eq:registry_size}, if we have $q$ qubits, then 
\begin{equation*}
    q  = n + n/d +2 \Rightarrow n = \floor*{ \frac{q-2}{1 + 1/d} }.
\end{equation*}
This implies that the number of $d$-length symbols in a string that can be handled by a QC is at best\footnote{QC architectural constraints may prevent the usage of all the hardware qubits.}
\begin{equation}\label{eq:z_estimate}
    z  \le \floor*{n/d} = \floor*{\frac{q-2}{d+1}}.
\end{equation}

\textbf{SE examples}: Based on Eq.~\ref{eq:z_estimate}, in the code coverage use-case, when $d=1$, we will be able to represent $559$ unique code blocks in a test case. 

In the trace comparison use-case case, the more unique trace points there are, the smaller the trace length that we can store (and vice versa). For example, suppose we have  $256$ unique trace points (consuming 8 bits per symbol, i.e., $d=8$). Then, the number of observations in a trace can go up to $124$. 

Thus, in 2023, we may be able to analyze small code bases. Given the exponential growth of QCs~\cite{tvede2020present}, we may expect to be able to analyze medium-to-large codebases by the end of the decade.

\textbf{BI examples}: In the DNA comparison use-case, $d=2$. Thus, we will be able to compare DNA sequence of length 373. In the mRNA use-case, assuming that all 64 codons are present, $d=6$. That is, we can represent mRNA sequence of length 159.

Given that the practical DNA comparison may require longer sequences, we may need to wait until the end of the decade or longer.

\section{Summary}
We introduced a quantum computing algorithm for calculating the Hamming distance for a string against a group of strings. The strings can have symbols drawn from an alphabet with an arbitrary number of entries. The algorithm requires only $n + n/d +2$ qubits for storage and retrieval.

We implement the algorithm using QisKit and encapsulate it in a Python class so that any Python programmer can readily leverage it. The code can be accessed via~\cite{dat:github}.

We show examples of leveraging the algorithm for two BI and two SE problems. The same principles can be applied to other use-cases from these and other domains. We conclude with an estimation of when the algorithm can be used for practical purposes.

\bibliography{references}


\begin{thebibliography}{35}

%%% the Web address.

\ifx \showCODEN    \undefined \def \showCODEN     #1{\unskip}     \fi
\ifx \showDOI      \undefined \def \showDOI       #1{#1}\fi
\ifx \showISBNx    \undefined \def \showISBNx     #1{\unskip}     \fi
\ifx \showISBNxiii \undefined \def \showISBNxiii  #1{\unskip}     \fi
\ifx \showISSN     \undefined \def \showISSN      #1{\unskip}     \fi
\ifx \showLCCN     \undefined \def \showLCCN      #1{\unskip}     \fi
\ifx \shownote     \undefined \def \shownote      #1{#1}          \fi
\ifx \showarticletitle \undefined \def \showarticletitle #1{#1}   \fi
\ifx \showURL      \undefined \def \showURL       {\relax}        \fi
% invisible to TeX
\providecommand\bibfield[2]{#2}
\providecommand\bibinfo[2]{#2}
\providecommand\natexlab[1]{#1}
\providecommand\showeprint[2][]{arXiv:#2}

\bibitem[\protect\citeauthoryear{??}{Aqu}{2021}]%
        {AquaAlgo12:online}
 \bibinfo{year}{2021}\natexlab{}.
\newblock \bibinfo{booktitle}{\emph{Aqua (Algorithms for QUantum Applications)
  (qiskit.aqua) — Qiskit 0.25.3 documentation}}.
\newblock
\urldef\tempurl%
\url{https://qiskit.org/documentation/apidoc/qiskit_aqua.html}
\showURL{%
\tempurl}


\bibitem[\protect\citeauthoryear{??}{Qis}{2021}]%
        {QiskitPartialMeasurement}
 \bibinfo{year}{2021}\natexlab{}.
\newblock \bibinfo{booktitle}{\emph{Mid-Circuit Measurements Tutorial}}.
\newblock
\urldef\tempurl%
\url{https://quantum-computing.ibm.com/lab/docs/iql/manage/systems/midcircuit-measurement/}
\showURL{%
\tempurl}


\bibitem[\protect\citeauthoryear{Abraham et~al\mbox{.}}{Abraham
  et~al\mbox{.}}{2019}]%
        {Qiskit}
\bibfield{author}{\bibinfo{person}{H{\'e}ctor Abraham} {et~al\mbox{.}}}
  \bibinfo{year}{2019}\natexlab{}.
\newblock \bibinfo{booktitle}{\emph{Qiskit: An Open-source Framework for
  Quantum Computing}}.
\newblock
\urldef\tempurl%
\url{https://doi.org/10.5281/zenodo.2562110}
\showDOI{\tempurl}


\bibitem[\protect\citeauthoryear{Bravo}{Bravo}{2018}]%
        {bravo2018calculating}
\bibfield{author}{\bibinfo{person}{Jos{\'e}~Manuel Bravo}.}
  \bibinfo{year}{2018}\natexlab{}.
\newblock \showarticletitle{Calculating Hamming distance with the IBM Q
  Experience}.
\newblock \bibinfo{journal}{\emph{Preprints}} (\bibinfo{year}{2018}).
\newblock
\urldef\tempurl%
\url{https://doi.org/10.20944/preprints201804.0164.v2}
\showDOI{\tempurl}


\bibitem[\protect\citeauthoryear{Cho}{Cho}{2020}]%
        {cho2020ibm}
\bibfield{author}{\bibinfo{person}{Adrian Cho}.}
  \bibinfo{year}{2020}\natexlab{}.
\newblock \showarticletitle{IBM promises 1000-qubit quantum computer---a
  milestone---by 2023}.
\newblock \bibinfo{journal}{\emph{Science}} (\bibinfo{year}{2020}).
\newblock
\urldef\tempurl%
\url{https://www.sciencemag.org/news/2020/09/ibm-promises-1000-qubit-quantum-computer-milestone-2023}
\showURL{%
\tempurl}


\bibitem[\protect\citeauthoryear{Davida, Frankel, Matt, and Peralta}{Davida
  et~al\mbox{.}}{1999}]%
        {davida1999relation}
\bibfield{author}{\bibinfo{person}{George~I Davida}, \bibinfo{person}{Y
  Frankel}, \bibinfo{person}{B Matt}, {and} \bibinfo{person}{R Peralta}.}
  \bibinfo{year}{1999}\natexlab{}.
\newblock \showarticletitle{On the relation of error correction and
  cryptography to an online biometric based identification scheme}. In
  \bibinfo{booktitle}{\emph{Workshop on coding and cryptography}}.
  \bibinfo{pages}{1--10}.
\newblock


\bibitem[\protect\citeauthoryear{Deaton, Murphy, Garzon, Franceschetti, and
  Jr.}{Deaton et~al\mbox{.}}{1996}]%
        {deaton1996good}
\bibfield{author}{\bibinfo{person}{Russell~J. Deaton}, \bibinfo{person}{R.~C.
  Murphy}, \bibinfo{person}{Max~H. Garzon}, \bibinfo{person}{Donald~R.
  Franceschetti}, {and} \bibinfo{person}{Stanley Edward~Stevens Jr.}}
  \bibinfo{year}{1996}\natexlab{}.
\newblock \showarticletitle{Good encodings for DNA-based solutions to
  combinatorial problems}. In \bibinfo{booktitle}{\emph{Proceedings of a
  {DIMACS} Workshop on {DNA} Based Computers}},
  \bibfield{editor}{\bibinfo{person}{Laura~F. Landweber} {and}
  \bibinfo{person}{Eric~B. Baum}} (Eds.), Vol.~\bibinfo{volume}{44}.
  \bibinfo{publisher}{{DIMACS/AMS}}, \bibinfo{pages}{247--258}.
\newblock


\bibitem[\protect\citeauthoryear{Doriguello and Montanaro}{Doriguello and
  Montanaro}{2019}]%
        {doriguello2019quantum}
\bibfield{author}{\bibinfo{person}{Jo{\~a}o~Fernando Doriguello} {and}
  \bibinfo{person}{Ashley Montanaro}.} \bibinfo{year}{2019}\natexlab{}.
\newblock \showarticletitle{Quantum sketching protocols for Hamming distance
  and beyond}.
\newblock \bibinfo{journal}{\emph{Physical Review A}} \bibinfo{volume}{99},
  \bibinfo{number}{6} (\bibinfo{year}{2019}), \bibinfo{pages}{062331}.
\newblock


\bibitem[\protect\citeauthoryear{Du, Wang, and Pang}{Du et~al\mbox{.}}{2004}]%
        {du2004anomaly}
\bibfield{author}{\bibinfo{person}{Ye Du}, \bibinfo{person}{Hui-Qiang Wang},
  {and} \bibinfo{person}{Yong-Gang Pang}.} \bibinfo{year}{2004}\natexlab{}.
\newblock \showarticletitle{An anomaly intrusion detection method using average
  Hamming distance}. In \bibinfo{booktitle}{\emph{Proceedings of 2004
  International Conference on Machine Learning and Cybernetics}},
  Vol.~\bibinfo{volume}{5}. \bibinfo{pages}{2914--2918}.
\newblock


\bibitem[\protect\citeauthoryear{Hamming}{Hamming}{1950}]%
        {hamming1950error}
\bibfield{author}{\bibinfo{person}{Richard~W. Hamming}.}
  \bibinfo{year}{1950}\natexlab{}.
\newblock \showarticletitle{Error Detecting and Error Correcting Codes}.
\newblock \bibinfo{journal}{\emph{Bell System Technical Journal}}
  \bibinfo{volume}{29}, \bibinfo{number}{2} (\bibinfo{year}{1950}),
  \bibinfo{pages}{147--160}.
\newblock


\bibitem[\protect\citeauthoryear{Horv{\'{a}}th, Gergely, Besz{\'{e}}des,
  Tengeri, Balogh, and Gyim{\'{o}}thy}{Horv{\'{a}}th et~al\mbox{.}}{2019}]%
        {horvath2019code}
\bibfield{author}{\bibinfo{person}{Ferenc Horv{\'{a}}th},
  \bibinfo{person}{Tam{\'{a}}s Gergely}, \bibinfo{person}{{\'{A}}rp{\'{a}}d
  Besz{\'{e}}des}, \bibinfo{person}{D{\'{a}}vid Tengeri},
  \bibinfo{person}{Gerg{\~{o}} Balogh}, {and} \bibinfo{person}{Tibor
  Gyim{\'{o}}thy}.} \bibinfo{year}{2019}\natexlab{}.
\newblock \showarticletitle{Code coverage differences of Java bytecode and
  source code instrumentation tools}.
\newblock \bibinfo{journal}{\emph{Softw. Qual. J.}} \bibinfo{volume}{27},
  \bibinfo{number}{1} (\bibinfo{year}{2019}), \bibinfo{pages}{79--123}.
\newblock


\bibitem[\protect\citeauthoryear{Hridoy, Ahmed, and Hosain}{Hridoy
  et~al\mbox{.}}{2015}]%
        {hridoy2015regression}
\bibfield{author}{\bibinfo{person}{Syed Akib~Anwar Hridoy},
  \bibinfo{person}{Faysal Ahmed}, {and} \bibinfo{person}{Md~Shazzad Hosain}.}
  \bibinfo{year}{2015}\natexlab{}.
\newblock \showarticletitle{Regression Testing based on Hamming Distance and
  Code Coverage}.
\newblock \bibinfo{journal}{\emph{Int. J. of Computer Applications}}
  \bibinfo{volume}{120}, \bibinfo{number}{14} (\bibinfo{year}{2015}),
  \bibinfo{pages}{1--5}.
\newblock


\bibitem[\protect\citeauthoryear{Kanuparthi, Rajendran, and Karri}{Kanuparthi
  et~al\mbox{.}}{2016}]%
        {kanuparthi2016controlling}
\bibfield{author}{\bibinfo{person}{Arun Kanuparthi},
  \bibinfo{person}{Jeyavijayan Rajendran}, {and} \bibinfo{person}{Ramesh
  Karri}.} \bibinfo{year}{2016}\natexlab{}.
\newblock \showarticletitle{Controlling your control flow graph}. In
  \bibinfo{booktitle}{\emph{2016 IEEE International Symposium on Hardware
  Oriented Security and Trust (HOST)}}. \bibinfo{pages}{43--48}.
\newblock


\bibitem[\protect\citeauthoryear{Kathuria, Ratan, McConnell, and
  Bekiranov}{Kathuria et~al\mbox{.}}{2020}]%
        {kathuria2020implementation}
\bibfield{author}{\bibinfo{person}{Kunal Kathuria}, \bibinfo{person}{Aakrosh
  Ratan}, \bibinfo{person}{Michael McConnell}, {and} \bibinfo{person}{Stefan
  Bekiranov}.} \bibinfo{year}{2020}\natexlab{}.
\newblock \showarticletitle{Implementation of a Hamming distance--like genomic
  quantum classifier using inner products on ibmqx2 and ibmq\_16\_melbourne}.
\newblock \bibinfo{journal}{\emph{Quantum machine intelligence}}
  \bibinfo{volume}{2}, \bibinfo{number}{1} (\bibinfo{year}{2020}),
  \bibinfo{pages}{1--26}.
\newblock


\bibitem[\protect\citeauthoryear{Khan and Miranskyy}{Khan and
  Miranskyy}{2021}]%
        {dat:github}
\bibfield{author}{\bibinfo{person}{Mushahid Khan} {and} \bibinfo{person}{Andriy
  Miranskyy}.} \bibinfo{year}{2021}\natexlab{}.
\newblock \bibinfo{booktitle}{\emph{QisKit implementation of the algorithm}}.
\newblock
\urldef\tempurl%
\url{https://github.com/miranska/qc-str}
\showURL{%
\tempurl}


\bibitem[\protect\citeauthoryear{Li, Ding, and Xie}{Li et~al\mbox{.}}{2019}]%
        {li2019tackling}
\bibfield{author}{\bibinfo{person}{Gushu Li}, \bibinfo{person}{Yufei Ding},
  {and} \bibinfo{person}{Yuan Xie}.} \bibinfo{year}{2019}\natexlab{}.
\newblock \showarticletitle{Tackling the Qubit Mapping Problem for NISQ-Era
  Quantum Devices}. In \bibinfo{booktitle}{\emph{Proceedings of the
  Twenty-Fourth International Conference on Architectural Support for
  Programming Languages and Operating Systems}} \emph{(\bibinfo{series}{ASPLOS
  '19})}. \bibinfo{pages}{1001–1014}.
\newblock


\bibitem[\protect\citeauthoryear{Malaiya}{Malaiya}{1995}]%
        {malaiya1995antirandom}
\bibfield{author}{\bibinfo{person}{Yashwant~K Malaiya}.}
  \bibinfo{year}{1995}\natexlab{}.
\newblock \showarticletitle{Antirandom testing: Getting the most out of
  black-box testing}. In \bibinfo{booktitle}{\emph{Proceedings of Sixth
  International Symposium on Software Reliability Engineering. ISSRE'95}}.
  IEEE, \bibinfo{pages}{86--95}.
\newblock


\bibitem[\protect\citeauthoryear{Miranskyy, Madhavji, Gittens, Davison,
  Wilding, and Godwin}{Miranskyy et~al\mbox{.}}{2007}]%
        {miranskyy2007iterative}
\bibfield{author}{\bibinfo{person}{Andriy~V Miranskyy},
  \bibinfo{person}{Nazim~H Madhavji}, \bibinfo{person}{Mechelle~S Gittens},
  \bibinfo{person}{Matthew Davison}, \bibinfo{person}{Mark Wilding}, {and}
  \bibinfo{person}{David Godwin}.} \bibinfo{year}{2007}\natexlab{}.
\newblock \showarticletitle{An iterative, multi-level, and scalable approach to
  comparing execution traces}. In \bibinfo{booktitle}{\emph{Proceedings of the
  the 6th joint meeting of the European software engineering conference and the
  ACM SIGSOFT symposium on The foundations of software engineering}}.
  \bibinfo{pages}{537--540}.
\newblock


\bibitem[\protect\citeauthoryear{Miranskyy, Madhavji, Gittens, Davison,
  Wilding, Godwin, and Taylor}{Miranskyy et~al\mbox{.}}{2008}]%
        {miranskyy2008sift}
\bibfield{author}{\bibinfo{person}{Andriy~V Miranskyy},
  \bibinfo{person}{Nazim~H Madhavji}, \bibinfo{person}{Mechelle~S Gittens},
  \bibinfo{person}{Matthew Davison}, \bibinfo{person}{Mark Wilding},
  \bibinfo{person}{David Godwin}, {and} \bibinfo{person}{Colin~A Taylor}.}
  \bibinfo{year}{2008}\natexlab{}.
\newblock \showarticletitle{SIFT: a scalable iterative-unfolding technique for
  filtering execution traces}. In \bibinfo{booktitle}{\emph{Proceedings of the
  2008 conference of the center for advanced studies on collaborative research:
  meeting of minds}}. \bibinfo{pages}{274--288}.
\newblock


\bibitem[\protect\citeauthoryear{Mohammadi-Kambs, H{\"o}lz, Somoza, and
  Ott}{Mohammadi-Kambs et~al\mbox{.}}{2017}]%
        {mohammadi2017hamming}
\bibfield{author}{\bibinfo{person}{Mina Mohammadi-Kambs},
  \bibinfo{person}{Kathrin H{\"o}lz}, \bibinfo{person}{Mark~M Somoza}, {and}
  \bibinfo{person}{Albrecht Ott}.} \bibinfo{year}{2017}\natexlab{}.
\newblock \showarticletitle{Hamming distance as a concept in DNA molecular
  recognition}.
\newblock \bibinfo{journal}{\emph{ACS omega}} \bibinfo{volume}{2},
  \bibinfo{number}{4} (\bibinfo{year}{2017}), \bibinfo{pages}{1302--1308}.
\newblock


\bibitem[\protect\citeauthoryear{Mount}{Mount}{2004}]%
        {mount2004bioinformatics}
\bibfield{author}{\bibinfo{person}{David~W Mount}.}
  \bibinfo{year}{2004}\natexlab{}.
\newblock \bibinfo{booktitle}{\emph{Bioinformatics: Sequence and Genome
  Analysis} (\bibinfo{edition}{2} ed.)}.
\newblock \bibinfo{publisher}{Cold Spring Harbor Laboratory Press}.
\newblock


\bibitem[\protect\citeauthoryear{Mrozek and Yarmolik}{Mrozek and
  Yarmolik}{2012}]%
        {mrozek2012antirandom}
\bibfield{author}{\bibinfo{person}{Ireneusz Mrozek} {and}
  \bibinfo{person}{Vyacheslav Yarmolik}.} \bibinfo{year}{2012}\natexlab{}.
\newblock \showarticletitle{Antirandom test vectors for BIST in
  hardware/software systems}.
\newblock \bibinfo{journal}{\emph{Fundamenta Informaticae}}
  \bibinfo{volume}{119}, \bibinfo{number}{2} (\bibinfo{year}{2012}),
  \bibinfo{pages}{163--185}.
\newblock


\bibitem[\protect\citeauthoryear{Nielsen and Chuang}{Nielsen and
  Chuang}{2010}]%
        {nielsen_chuang_2010}
\bibfield{author}{\bibinfo{person}{Michael~A. Nielsen} {and}
  \bibinfo{person}{Isaac~L. Chuang}.} \bibinfo{year}{2010}\natexlab{}.
\newblock \bibinfo{booktitle}{\emph{Quantum Computation and Quantum
  Information: 10th Anniversary Edition}}.
\newblock \bibinfo{publisher}{Cambridge Univ. Press}.
\newblock


\bibitem[\protect\citeauthoryear{Preskill}{Preskill}{2018}]%
        {Preskill2018quantumcomputing}
\bibfield{author}{\bibinfo{person}{John Preskill}.}
  \bibinfo{year}{2018}\natexlab{}.
\newblock \showarticletitle{Quantum {C}omputing in the {NISQ} era and beyond}.
\newblock \bibinfo{journal}{\emph{{Quantum}}}  \bibinfo{volume}{2}
  (\bibinfo{date}{Aug.} \bibinfo{year}{2018}), \bibinfo{pages}{79}.
\newblock
\showISSN{2521-327X}
\urldef\tempurl%
\url{https://doi.org/10.22331/q-2018-08-06-79}
\showDOI{\tempurl}


\bibitem[\protect\citeauthoryear{Sarkar, Al-Ars, Almudever, and Bertels}{Sarkar
  et~al\mbox{.}}{2019}]%
        {sarkar2019algorithm}
\bibfield{author}{\bibinfo{person}{Aritra Sarkar}, \bibinfo{person}{Zaid
  Al-Ars}, \bibinfo{person}{Carmen~G. Almudever}, {and} \bibinfo{person}{Koen
  Bertels}.} \bibinfo{year}{2019}\natexlab{}.
\newblock \bibinfo{title}{An algorithm for DNA read alignment on quantum
  accelerators}.
\newblock
\newblock
\showeprint[arxiv]{1909.05563}~[quant-ph]


\bibitem[\protect\citeauthoryear{Shan, Zhang, and He}{Shan
  et~al\mbox{.}}{2017}]%
        {shan2017machine}
\bibfield{author}{\bibinfo{person}{Weiwei Shan}, \bibinfo{person}{Shuai Zhang},
  {and} \bibinfo{person}{Yukun He}.} \bibinfo{year}{2017}\natexlab{}.
\newblock \showarticletitle{Machine learning based side-channel-attack
  countermeasure with hamming-distance redistribution and its application on
  advanced encryption standard}.
\newblock \bibinfo{journal}{\emph{Electronics Letters}} \bibinfo{volume}{53},
  \bibinfo{number}{14} (\bibinfo{year}{2017}), \bibinfo{pages}{926--928}.
\newblock


\bibitem[\protect\citeauthoryear{Siraichi, Santos, Collange, and
  Pereira}{Siraichi et~al\mbox{.}}{2018}]%
        {10.1145/3168822}
\bibfield{author}{\bibinfo{person}{Marcos~Yukio Siraichi},
  \bibinfo{person}{Vin\'{\i}cius Fernandes~dos Santos},
  \bibinfo{person}{Sylvain Collange}, {and} \bibinfo{person}{Fernando
  Magno~Quintao Pereira}.} \bibinfo{year}{2018}\natexlab{}.
\newblock \showarticletitle{Qubit Allocation}. In
  \bibinfo{booktitle}{\emph{Proceedings of the 2018 International Symposium on
  Code Generation and Optimization}} \emph{(\bibinfo{series}{CGO 2018})}.
  \bibinfo{pages}{113–125}.
\newblock


\bibitem[\protect\citeauthoryear{Sousa, dos Santos, de~Veras, de~Oliveira, and
  da~Silva}{Sousa et~al\mbox{.}}{2020}]%
        {sousa2020parametric}
\bibfield{author}{\bibinfo{person}{Rodrigo~S. Sousa}, \bibinfo{person}{Priscila
  G.~M. dos Santos}, \bibinfo{person}{Tiago Mendon{\c{c}}a~Lucena de Veras},
  \bibinfo{person}{Wilson~Rosa de Oliveira}, {and}
  \bibinfo{person}{Adenilton~J. da Silva}.} \bibinfo{year}{2020}\natexlab{}.
\newblock \showarticletitle{Parametric Probabilistic Quantum Memory}.
\newblock \bibinfo{journal}{\emph{Neurocomputing}}  \bibinfo{volume}{416}
  (\bibinfo{year}{2020}), \bibinfo{pages}{360--369}.
\newblock


\bibitem[\protect\citeauthoryear{Stabili, Marchetti, and Colajanni}{Stabili
  et~al\mbox{.}}{2017}]%
        {dario2017detecting}
\bibfield{author}{\bibinfo{person}{Dario Stabili}, \bibinfo{person}{Mirco
  Marchetti}, {and} \bibinfo{person}{Michele Colajanni}.}
  \bibinfo{year}{2017}\natexlab{}.
\newblock \showarticletitle{Detecting attacks to internal vehicle networks
  through Hamming distance}. In \bibinfo{booktitle}{\emph{2017 AEIT
  International Annual Conference}}. \bibinfo{pages}{1--6}.
\newblock


\bibitem[\protect\citeauthoryear{Taheri, Ghahramani, Javidan, Shojafar,
  Pooranian, and Conti}{Taheri et~al\mbox{.}}{2020}]%
        {taheri2020similarity}
\bibfield{author}{\bibinfo{person}{Rahim Taheri}, \bibinfo{person}{Meysam
  Ghahramani}, \bibinfo{person}{Reza Javidan}, \bibinfo{person}{Mohammad
  Shojafar}, \bibinfo{person}{Zahra Pooranian}, {and} \bibinfo{person}{Mauro
  Conti}.} \bibinfo{year}{2020}\natexlab{}.
\newblock \showarticletitle{Similarity-based Android malware detection using
  Hamming distance of static binary features}.
\newblock \bibinfo{journal}{\emph{Future Gener. Comput. Syst.}}
  \bibinfo{volume}{105} (\bibinfo{year}{2020}), \bibinfo{pages}{230--247}.
\newblock


\bibitem[\protect\citeauthoryear{Tang, Yu, Aref, Malluhi, and Ouzzani}{Tang
  et~al\mbox{.}}{2015}]%
        {tang2015efficient}
\bibfield{author}{\bibinfo{person}{MingJie Tang}, \bibinfo{person}{Yongyang
  Yu}, \bibinfo{person}{Walid~G Aref}, \bibinfo{person}{Qutaibah~M Malluhi},
  {and} \bibinfo{person}{Mourad Ouzzani}.} \bibinfo{year}{2015}\natexlab{}.
\newblock \showarticletitle{Efficient Processing of Hamming-Distance-Based
  Similarity-Search Queries Over MapReduce}. In
  \bibinfo{booktitle}{\emph{EDBT}}. \bibinfo{pages}{361--372}.
\newblock


\bibitem[\protect\citeauthoryear{Trugenberger}{Trugenberger}{2001}]%
        {pqmct}
\bibfield{author}{\bibinfo{person}{Carlo Trugenberger}.}
  \bibinfo{year}{2001}\natexlab{}.
\newblock \showarticletitle{Probabilistic Quantum Memories}.
\newblock \bibinfo{journal}{\emph{Physical review letters}}
  \bibinfo{volume}{87} (\bibinfo{date}{09} \bibinfo{year}{2001}),
  \bibinfo{pages}{067901}.
\newblock


\bibitem[\protect\citeauthoryear{Tvede}{Tvede}{2020}]%
        {tvede2020present}
\bibfield{author}{\bibinfo{person}{Lars Tvede}.}
  \bibinfo{year}{2020}\natexlab{}.
\newblock \showarticletitle{The Present And Future Of Quantum Computing
  Expansion}.
\newblock \bibinfo{journal}{\emph{Forbes}} (\bibinfo{year}{2020}).
\newblock
\urldef\tempurl%
\url{https://www.forbes.com/sites/forbesbusinesscouncil/2020/07/14/the-present-and-future-of-quantum-computing-expansion/?sh=4815b6f943b9}
\showURL{%
\tempurl}


\bibitem[\protect\citeauthoryear{Wu, Jandhyala, Malaiya, and Jayasumana}{Wu
  et~al\mbox{.}}{2008}]%
        {wu2008antirandom}
\bibfield{author}{\bibinfo{person}{Shen~Hui Wu}, \bibinfo{person}{Sridhar
  Jandhyala}, \bibinfo{person}{Yashwant~K Malaiya}, {and}
  \bibinfo{person}{Anura~P Jayasumana}.} \bibinfo{year}{2008}\natexlab{}.
\newblock \showarticletitle{Antirandom Testing: A Distance-Based Approach.}
\newblock \bibinfo{journal}{\emph{VLSI Design}} \bibinfo{volume}{2008},
  \bibinfo{number}{2} (\bibinfo{year}{2008}), \bibinfo{pages}{1--9}.
\newblock


\bibitem[\protect\citeauthoryear{Xie, Qiu, and Cai}{Xie et~al\mbox{.}}{2018}]%
        {hamdistancebooleanfunctions2018}
\bibfield{author}{\bibinfo{person}{Zhengwei Xie}, \bibinfo{person}{Daowen Qiu},
  {and} \bibinfo{person}{Guangya Cai}.} \bibinfo{year}{2018}\natexlab{}.
\newblock \showarticletitle{Quantum algorithms on Walsh transform and Hamming
  distance for Boolean functions}.
\newblock \bibinfo{journal}{\emph{Quantum Information Processing}}
  \bibinfo{volume}{17}, \bibinfo{number}{6} (\bibinfo{year}{2018}),
  \bibinfo{pages}{1--17}.
\newblock


\end{thebibliography}

\end{document}